\documentclass[11pt,dvips]{article}
\usepackage{epsfig,times} 
\usepackage{wrapfig}
\makeatletter
\renewcommand{\section}{\@startsection{section}{1}{0in}
	{0.4\baselineskip}{0.1\baselineskip}{\Large\bf}}
\renewcommand{\subsection}{\@startsection{subsection}{2}{0in}
	{0.25\baselineskip}{-\baselineskip}{\large\bf}}
\renewcommand{\subsubsection}{\@startsection{subsubsection}{3}{0in}
	{0.1\baselineskip}{-\baselineskip}{\normalsize\bf}}
\makeatother
\def\arcdeg{\hbox{$^\circ$}}

\def\fdg{\hbox{$.\!\!^\circ$}}

\newcommand{\gtapprox}{\hbox{\raise0.5ex \hbox{$>$}
   \kern-1.1em \lower0.5ex \hbox{$\sim$}}}

\newcommand{\ltapprox}{\hbox{\raise0.5ex \hbox{$<$}
   \kern-1.1em \lower0.5ex \hbox{$\sim$}}}
\begin{document}

%
\thispagestyle{myheadings}
%
\markright{OG 2.2.12}
\begin{center}
%
{\LARGE \bf 
Evidence of 10--100~TeV Electrons in Supernova Remnants
}
\end{center}

\begin{center}
%
%
{\bf G.E. Allen$^{1}$, E.V. Gotthelf$^{2}$, R. Petre$^{1}$}\\
{\it $^{1}$NASA/Goddard Space Flight Center, LHEA, Code 661, Greenbelt, MD
20771, USA\\
$^{2}$Columbia Astrophysics Laboratory, Pupin Hall, Columbia University,
550 West 120th Street, New York, NY 10027, USA}
\end{center}

\begin{center}
{\large \bf Abstract\\}
\end{center}
\vspace{-0.5ex}
%
%
Analyses of the X-ray data of the five young shell-type supernova remnants 
Cas~A, Kepler, Tycho, SN~1006, and RCW~86 suggest that some of the X-ray 
emission of these sources is non-thermal.  
This non-thermal emission is qualitatively consistent with models of the 
broad-band (radio-to-X-ray) synchrotron spectra of remnants and does not 
seem to be consistent with other non-thermal X-ray emission processes.  
If this emission is produced by synchrotron radiation, the radio-to-X-ray 
synchrotron spectra imply that the electron spectra have differential 
spectral indices of about 2.2 and exponential cut offs at energies $\sim 
10$~TeV.  
If the remnants also accelerate cosmic-ray nuclei, the total energies of the 
cosmic rays in the remnants are estimated to be $\sim 1$--$5 \times 10^{49}$ 
erg.  
Therefore, the shapes of the cosmic-ray electron spectra, the maximum energies
of the cosmic-ray electrons, and the total cosmic-ray energies of the five 
young remnants seem to be consistent with the idea that Galactic cosmic rays 
are predominantly accelerated in the shocks of supernova remnants.
%

\vspace{1ex}

%
%
\section{Introduction:}
\label{int}

For many years the X-ray emission of shell-type (as opposed to plerionic) 
supernova remnants was modeled in terms of only the emission of hot thermal 
plasmas.  
However, it has recently become clear that at least some young shell-type 
supernova remnants (SN~1006, Koyama et~al.\ 1995, Allen et~al.\ 1999; Cas~A, 
Allen et~al.\ 1997; G347.3$-$0.5, Koyama et~al.\ 1997; IC~443, Keohane
et~al.\ 1997) produce non-thermal emission as well.  
At least for SN~1006 and Cas~A, the only plausible description of the 
non-thermal X-ray emission is synchrotron emission by 10--100~TeV electrons 
(Allen et~al.\ 1999; Allen et~al.\ 1997).
The non-thermal X-ray spectra of these two remnants are qualitatively
consistent with simple models of the radio-to-X-ray synchrotron spectra, but
are inconsistent with the predicted shapes and flux levels of other processes
such as bremsstrahlung emission or inverse Compton scattering of the cosmic 
microwave background radiation (fig.~\ref{fig1}).
Furthermore, the high-energy (i.e.\ non-thermal--dominated) X-ray images
differ
substantially from lower-energy (thermal-dominated) X-ray images 
(Willingale et~al.\ 1996; Holt et~al.\ 1994; Vink et~al.\ 1999), but are 
similar to the radio synchrotron images of the remnants (Reynolds and 
Gilmore 1986; Anderson and Rudnick 1995).  
These two clues provide very-strong support for the idea that the non-thermal
X-ray emission of SN~1006 and Cas~A is synchrotron radiation.
The presence of very-high--energy electrons in SN~1006 is corroborated by the
detection of TeV gamma-ray emission from this remnant (Tanimori et~al.\ 1998)

These results have very important implications for the study of Galactic 
cosmic-ray acceleration.
Galactic cosmic rays, up to an energy of about 3000~TeV (the ``knee''),
are thought to be predominantly accelerated in the shocks of supernova 
remnants.  
Since the shock-acceleration process of supernova remnants depends on the 
magnetic rigidity of the particles, it may be the case that the cosmic-ray 
particles at 3000~TeV are principally iron and that protons (and electrons) 
are only accelerated to energies of about 100~TeV (Lagage and Cesarsky 1983).
However, it has been difficult to confirm that supernova remnants accelerate
particles to such high energies.
Although previous radio and gamma-ray observations of remnants reveal
evidence of non-thermal particles in many supernova remnants, these results
are limited to particle energies (\ltapprox\ 10~GeV) that are well below 
100~TeV (except for the TeV gamma-ray results of Tanimori et~al.\ 1998).
The non-thermal X-ray data of SN~1006 provided the first evidence that 
supernova remnants accelerate particles to very-high energies (Koyama et~al.\
1995; Reynolds 1996).
Subsequent observations reveal that such X-ray emission is not unique to
SN~1006 (Allen et~al.\ 1997; Koyama et~al.\ 1997; Keohane et~al.\ 1997).
Simple models of the radio-to-X-ray synchrotron spectra of supernova remnants
(Reynolds 1998) have been used to show that the electron spectra of SN~1006
and Cas~A have exponential cut offs with e-folding energies $\epsilon 
\sim 20$ and 5~TeV, respectively.
Therefore, the study of the non-thermal X-ray emission of young shell-type
supernova remnants provides a powerful means of studying the acceleration of
the highest-energy cosmic-ray electrons in situ.

\section{Data and Analyses:}
\label{dat}

Several supernova remnants have been observed using the Proportional Counter
Array (PCA) on the Rossi X-Ray Timing Explorer satellite.
The PCA is a spectrophotometer comprised of an array of five co-aligned
proportional counter units that are mechanically collimated to have a
field-of-view of $1\arcdeg$~FWHM (Jahoda et~al.\ 1996).
We have analyzed PCA data for the five, young, shell-type remnants Cas~A, 
Kepler, Tycho, SN~1006, and RCW~86.
%
%
%
%
%
These data were screened to exclude the time intervals during which (1) one 
or more of the five proportional counter units was off, (2) the elevation of
the source above the limb of the Earth $< 10^{\circ}$, (3) the PCA 
background model is not well behaved, and (4) the nominal pointing direction 
of the PCA $> 0\fdg02$ from the specified direction of the source.
After applying these selection criteria, 169, 44, 81, 18, and 32~ks of data 
were used to construct the X-ray spectra of Cas~A, Kepler, Tycho, SN~1006, 
and RCW~86, respectively.

The results of the spectral analyses are shown in figure~\ref{fig3}.
The spectral data below 10~keV for Cas~A, Kepler, Tycho, and RCW~86 and below
7~keV for SN~1006 are excluded to insure that the spectra of the remnants 
are dominated by non-thermal emission.

\section{Discussion and Conclusion:}
\label{dis}

At energies above 10~keV, non-thermal X-ray emission dominates the spectra of 
both SN~1006 and Cas~A.
Both of these spectra can be described by power laws with photon indices of 
$\Gamma = 3.0 \pm 0.2$ (Allen et~al.\ 1997; Allen et~al.\ 1999).
As shown in figure~\ref{fig3}, similar results are obtained for the remnants 
Kepler ($\Gamma = 3.0 \pm 0.2$), Tycho ($\Gamma = 3.2 \pm 0.1$), and RCW~86 
($\Gamma = 3.3 \pm 0.2$).
Therefore, the high-energy non-thermal X-ray spectra of the five remnants may
be produced by a common emission mechanism.
Since the high-energy non-thermal X-ray spectra of SN~1006 and Cas~A are
produced by synchrotron radiation from 10--100~TeV electrons, the results 
shown in figure~\ref{fig3} support the conclusion that all young shell-type 
supernova remnants accelerate electrons to very--high-energies.  
If more detailed analyses of the X-ray emission of Kepler, Tycho, RCW~86, and
other young shell-type supernova remnants confirm this conclusion, the results
will have very important implications for the origins of Galactic cosmic rays.

\begin{wrapfigure}{r}{3.5in}
\epsfxsize=3.5in
\centerline{\epsffile{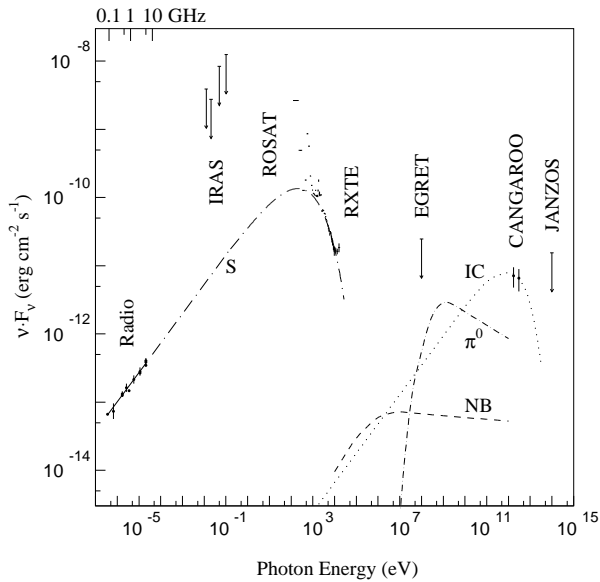}}
\caption{
The multi-wavelength photon spectrum of SN~1006.
The four broken curves are estimates of the fluxes from synchrotron 
radiation (S), non-thermal bremsstrahlung emission (NB), inverse Compton 
scattering of the cosmic microwave background radiation (IC), and the decay 
of neutral pions ($\pi^{0}$).
The details about the spectral data and the estimates of the photon spectra 
are described in detail by Allen et~al.\ (1999).  
In contrast to the other emission processes, the synchrotron spectral model is
consistent with both the shape and the flux level of the high-energy X-ray 
spectrum of SN~1006.
\label{fig1}}
\end{wrapfigure}

\begin{wrapfigure}{r}{3.5in}
\epsfxsize=3.5in
\centerline{\epsffile{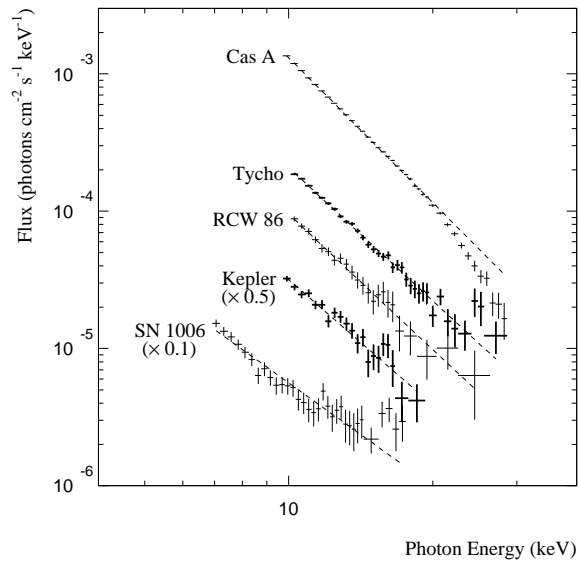}}
\caption{
The PCA high-energy X-ray spectra of the young shell-type supernova remnants
Cas~A, Tycho, RCW~86, Kepler, and SN~1006.
The spectra of Kepler and SN~1006 have been multiplied by factors of 0.5 and
0.1, respectively, to help distinguish one spectrum from the others.
The high-energy X-ray spectra of the remnants are non-thermal and have very 
similar spectral shapes.
Since the high-energy non-thermal emission of Cas~A and SN~1006 are produced 
by X-ray synchrotron emission (Allen et~al. 1997; Allen et~al.\ 1999), it
appears that the high-energy X-ray spectra of Kepler, Tycho, and RCW~86 may 
also be produced by X-ray synchrotron emission.  
\label{fig3}}
\end{wrapfigure}

The energy at which the radio-to-X-ray synchrotron spectra roll over (see
fig.~\ref{fig1}) can be used to determine the exponential cut-off energy of 
the electron spectra of the remnants.
For example, if the magnetic field strength of SN~1006 $\sim 10$~$\mu$G 
(Tanimori et~al.\ 1998) and the energy of the roll off in the synchrotron 
spectrum $\sim 100$~eV (fig.~\ref{fig1}), the e-folding energy of the cut 
off in the electron spectrum of SN~1006 $\epsilon \sim 20$~TeV.  
Estimates of the e-folding energies of the cut offs in the electron spectra 
of the other remnants yield similar results ($\epsilon \sim 5$, 20, 10, and 
10~TeV for Cas~A, Kepler, Tycho, and RCW~86, respectively).
Since these results are sensitive to the assumed shape of the electron
spectrum
and the strength of the magnetic field, they should be regarded as
order-of-magnitude estimates.
Nevertheless, it is interesting that all five of the estimates are below
100~TeV.
If the estimates are accurate, the results may indicate that the cosmic-ray 
electrons in the five remnants have not yet reached their maximum energy, 
that these remnants do not accelerate cosmic-ray electrons to energies above
$\sim 10$~TeV, or that the maximum energy of the electrons (but not the 
nuclei) is regulated by radiative losses.
More work is needed to differentiate between these possibilities.

Two other clues support the idea that Galactic cosmic rays are accelerated in
the shocks of supernova remnants.
One clue is the typical electron spectral index of the young remnants of
figure~\ref{fig3}.
A review of radio spectral indices of the five remnants ($\alpha = 0.77$,
0.64, 0.61, 0.57, and 0.6 for Cas~A, Kepler, Tycho, SN~1006, and RCW~86,
respectively, Green 1998) reveals that the typical differential spectral 
index of the electrons producing the radio spectra is about $\Gamma = 2.2$ 
($= 2 \alpha + 1$).
This index is consistent with the index expected for cosmic-ray
accelerators because the observed differential spectral index of the 
cosmic-ray protons observed at Earth ($\Gamma = 2.8$, Asakimori et~al.\ 1998) 
less the inferred spectral steepening due to an energy-dependent escape of 
the cosmic rays from the Galaxy ($\Delta\Gamma = 0.6$, Swordy et~al.\ 1990) 
is about 2.2.

%
%
The second clue also stems from an analysis of the radio spectral data.
The radio data of a remnant may be used to estimate the total energy of the 
cosmic-ray particles in the remnant.
Such estimates depend on some assumptions about the strengths of the 
magnetic fields and about the cosmic-ray electron and proton spectra of the 
remnants.
For simplicity, we assume (1) that the magnetic fields of the remnants are
$B = 10^{3}$, $10^{2}$, $10^{2}$, $10$, and $10^{2}$, for Cas~A, Kepler,
Tycho, SN~1006, and RCW~86, respectively, (2) that the 
cosmic-ray electron and proton spectra are described by $dN_{\rm e,p}/dE 
\propto (E + m_{\rm e,p}c^{2})(E^{2} + 2m_{\rm e,p}c^{2}E)^{-(\Gamma + 1)/2} 
e^{-E/\epsilon}$ (Bell 1978), (3) that the high-energy spectral indices 
($\Gamma$ ($= 2\alpha + 1$) of the electrons and protons are the same, (4) 
that the e-folding energy of the exponential cut offs $\epsilon = 10$~TeV, 
and (5) that non-thermal electrons outnumber non-thermal protons by a factor 
of 1.2.  
The last assumption follows from the assumptions that the relative elemental
abundances of the cosmic rays are comparable to the relative elemental
abundances of the solar system and that all hydrogen and helium nuclei are
fully ionized.
This set of assumptions naturally leads to the result that cosmic-ray protons
outnumber cosmic-ray electrons by a factor $\sim 100$ at 1~GeV, as is observed
at Earth.
Using these assumptions, we find that, at the present, the total energies of 
the cosmic-ray particles in the remnants $U_{\rm cr} \sim 5$, 2, 1, 2, and
$1 \times 10^{49}$~erg for Cas~A, Kepler, Tycho, SN~1006, and RCW~86,
respectively.
These order-of-magnitude estimates are comparable to estimates of the total 
energy that is needed, on average, per remnant, over their lifetimes ($\sim 
3$--$10 \times 10^{49}$~erg, Blandford and Eichler 1987; Lingenfelter 1992).

In summary, analyses of the high-energy X-ray spectra of five young supernova
remnants suggest that the remnants have similar non-thermal spectral 
properties at energies $\gtapprox\ 10$~keV.
These spectra are qualitatively consistent with models of the radio-to-X-ray
synchrotron spectra of supernova remnants.
If this emission is produced by synchrotron radiation, the radio-to-X-ray 
synchrotron spectra imply that the electron spectra have differential 
spectral indices of about 2.2 and exponential cut offs at energies $\sim 
10$~TeV.  
If the remnants also accelerate cosmic-ray nuclei, the total energies of the 
cosmic rays in the remnants are estimated to be $\sim 1$--$5 \times 10^{49}$ 
erg.  
These results appear to be consistent with the idea that Galactic cosmic-ray
electrons (and, presumably, nuclei) are predominantly accelerated in the
shocks of supernova remnants.

%
%
%
%
%
%
\vspace{1ex}
\begin{center}
{\Large\bf References}
\end{center}
%
%
%

\noindent
Allen, G.\ E.\ et~al.\ 1997, ApJ, 487, L97 \\
Allen, G.\ E.\ et~al.\ 1999, In preparation \\
Anderson, M.\ C., \& Rudnick, L.\ 1995, ApJ, \\ 
   \hspace*{0.25in} 441, 307 \\
Asakimori, K.\ et~al.\ 1998, ApJ, 502, 278 \\
Bell, A.\ R.\ 1978, MNRAS, 182, 443 \\
Blandford, R.\ \& Eichler, D.\ 1987, Phys. \\
   \hspace*{0.25in} Rep., 154, 1 \\
Holt, S.\ S., et~al.\ 1994, PASJ, 46, L151 \\
Jahoda, K., et~al.\ 1996, in EUV, X-ray and \\
   \hspace*{0.25in} Gamma-ray Instrumentation for Space \\
   \hspace*{0.25in} Astronomy VII, ed. Siegmund, O.\ H.\ \\
   \hspace*{0.25in} W, \& Grummin, M.\ A.\ Proc.\ SPIE, \\
   \hspace*{0.25in} 2808, 59 \\
Keohane, J.\ W.\ et~al.\ 1997, ApJ 484, 350 \\
Koyama, K.\ et~al.\ 1995, Nature, 378, 255 \\
Koyama, K.\ et~al.\ 1997, PASJ, 49, L7 \\
Lagage,~P.~O., \& Cesarsky, C.~J.\ 1983, A\&A, \\
   \hspace*{0.25in} 125, 249 \\
Lingenfelter, R.\ E.\ 1992, In {The Astronomy \\
   \hspace*{0.25in} and Astrophysics Encyclopedia}, Edited \\
   \hspace*{0.25in} by Maran, S.\ 139, Van Nostrand Rein- \\
   \hspace*{0.25in} hold Publishers \\
Reynolds, S.\ P.\ 1996, ApJ, 459, L13 \\
Reynolds, S.\ P.\ 1998, ApJ, 493, 375 \\
Reynolds,~S.~P., \& Gilmore, D.\ M.\ 1986, AJ, \\
   \hspace*{0.25in} 92, 1138 \\
Swordy, S.\ P.\ et~al.\ 1990, ApJ, 349, 625 \\
Tanimori, T.\ et~al.\ 1998, ApJ, 497, L25 \\
The, L.-S., et~al.\ 1996, A\&ApS, 120, 357 \\
Vink, J.\ et~al.\ 1999, A\&A, In press \\
Willingale, R., et~al.\ 1996, MNRAS, 278, 749 \\

\end{document}